# Shortest Paths in Microseconds

Rachit Agarwal
University of Illinois
Urbana-Champaign, USA
agarwa16@illinois.edu

Matthew Caesar
University of Illinois
Urbana-Champaign, USA
caesar@illinois.edu

P. Brighten Godfrey
University of Illinois
Urbana-Champaign, USA
pbg@illinois.edu

Ben Y. Zhao
University of California
Santa Barbara, USA
ravenben@cs.ucsb.edu

## ABSTRACT

Computing *shortest paths* is a fundamental primitive for several social network applications including socially-sensitive ranking, location-aware search, social auctions and social network privacy. Since these applications compute paths in response to a user query, the goal is to minimize latency while maintaining feasible memory requirements. We present ASAP, a system that achieves this goal by exploiting the structure of social networks.

ASAP preprocesses a given network to compute and store a *partial shortest path tree* (PSPT) for each node. The PSPTs have the property that for any two nodes, each edge along the shortest path is with high probability contained in the PSPT of at least one of the nodes. We show that the structure of social networks enable the PSPT of each node to be an extremely small fraction of the entire network; hence, PSPTs can be stored efficiently and each shortest path can be computed extremely quickly.

For a real network with 5 million nodes and 69 million edges, ASAP computes a shortest path for most node pairs in less than 49 *microseconds* per pair. ASAP, unlike any previous technique, also computes hundreds of paths (along with corresponding distances) between any node pair in less than 100 *microseconds*. Finally, ASAP admits efficient implementation on distributed programming frameworks like MapReduce.

## 1. INTRODUCTION

Computing distances and paths is a fundamental primitive in social network analysis — in LinkedIn, it is desirable to compute short paths from a job seeker to a potential employer; in social auction sites, distances and paths are used to identify trustworthy sellers [22]; and distances are used to compute rankings in social search [18, 25]. In addition, applications like socially-sensitive and location-aware search [5,20] require computing paths between a user and content of potential interest to the user. These applications require or can benefit from computing *shortest paths for most queries*, which we focus on in this work.

This paper is particularly motivated by the following two applications. Consider a professional social network (LinkedIn, Microsoft Academic Search, etc.) where a user X search for another user Y. Social networks desire to provide the user X a list of possible paths to the user Y, with each path ranked according to some metric that depends on the length of the path. The problem here is to *quickly* compute multiple "short" paths (and corresponding path lengths) between a pair of users. The second application relates to social search [25] and socially-sensitive search ranking. Here, a social network user X searches for user Y (or, for content Y); however, multiple social network users (or contents) may satisfy the search criteria. Social networks desire to output a list of the search results ranked according to the distance between user X and each user that satisfies the search criteria for Y. The problem here is to quickly compute the distance (and path) between a user X and multiple users (or contents).

Scalable computation of shortest paths on social networks is challenging for two reasons. First, applications above compute paths in response to a user query and hence, have rather stringent latency requirements [16]. This precludes the obvious option of running a shortest path algorithm like $A^\star$ search [9, 10] or bidirectional search [10] for each query — as we will show in §2, these algorithms require hundreds of milliseconds even on moderate size networks.

Second, the massive size of social networks make it infeasible to precompute and store shortest paths; even for a social network with 3 million users, this would require 4.5 trillion entries. Citing lack of efficient techniques for computing shortest paths, a number of papers have developed techniques to compute approxi-mate distances and paths [11,18,19,23,28]. We delay a complete discussion of related work to §5; however, we note that these techniques either compute paths that are significantly longer than the actual shortest path or do not meet the latency requirements.

We present ASAP, a system that quickly computes shortest paths for most queries on social networks while maintaining feasible memory requirements. ASAP preprocesses the network to compute a *partial shortest path tree* (PSPT) for each node. PSPTs have the property that for any two nodes $s, t$, each edge along the shortest path is very highly likely to be contained in the PSPT of either $s$ or $t$; that is, there is one node $w$ that belongs to the PSPT of both $s$ and $t$. Hence, a shortest path can be computed by combining paths $s \rightsquigarrow w$ and $t \rightsquigarrow w$. For the unlikely case of PSPTs not intersecting along the shortest path, ASAP computes a path that is at most one hop longer than the shortest path.

ASAP presents several contributions. First, it focuses on a much harder problem of computing *shortest paths for most queries* and even on networks with millions of nodes and edges, computes shortest paths in *tens of microseconds*. Second, ASAP demonstrates and exploits the observation that the structure of social networks enable the PSPT of each node to be an extremely small fraction of the entire network; hence, PSPTs can be stored efficiently and each shortest path can be computed extremely quickly. It is known that planar graphs exhibit a similar structure [6], but that social networks exhibit such a structure despite having significantly different properties is interesting in its own right. Finally, unlike most previous works, ASAP admits efficient distributed implementation and can be easily mapped on distributed programming frameworks like MapReduce.

ASAP, for the LiveJournal network (roughly 5 million nodes and 69 million edges), computes the shortest path between 99.83% of the node pairs in less than 49 *microseconds* — 3196× faster than the bidirectional shortest path algorithm [10]; computes a path that is at most one hop longer than the shortest path[1] for an additional 0.15% of the node pairs in 49 microseconds; and runs a bidirectional shortest path algorithm for the remaining 0.02% of the node pairs[2]. These results enable the second class of applications discussed earlier. For the first set of applications, ASAP allows computing hundreds of paths and corresponding distances between more than 99.98% of the node pairs in less than 100 microseconds without any change in the data structure for single shortest path computation, thus enabling distance-based social search and ranking in a unified way.

---
[1] These are the cases when the PSPTs of the node pair intersect, but not along the shortest path.
[2] These are the cases when the PSPTs of the node pair do not intersect.

## 2. ASAP

We start the section by formally defining the PSPT of a node, a structure that forms the most basic component of ASAP (§2.1). We then describe the ASAP algorithm for computing the shortest path between a given pair of nodes (§2.2). In §2.3, we describe a low-memory, low-latency implementation of ASAP. Finally, we discuss extensions of ASAP that allow computing multiple paths and further speeding up ASAP for the special case of unweighted graphs (§2.4). We assume that the input network $G = (V, E)$ is an undirected weighted network; each edge is assigned a non-negative weight and each node is assigned a unique identifier.

### 2.1 Partial Shortest Path Trees

We now define the PSPT of each node. At a high level, we will require that the PSPTs of any pair of nodes $s, t$ satisfy the following property: there exists a node $w$ along the shortest path between $s$ and $t$ such that (1) $w$ is contained in the PSPT of both $s$ and $t$ (or equivalently, the two PSPTs *intersect* along the shortest path); (2) the path $s \rightsquigarrow w$ is contained in the PSPT of $s$; and (3) the path $t \rightsquigarrow w$ is contained in the PSPT of $w$.

To start with, note that nodes that have only one neighbor (or equivalently, degree-1 nodes) can never lie along any shortest path; hence, PSPTs do not need to contain degree-1 nodes. To this end, let $G' = (V', E')$ be the network achieved by removing from $V$ all degree-1 nodes and from $E$ all edges incident on degree-1 nodes. Then, the PSPT of size $\beta$ of any node $u$ is the set of $\beta$ closest nodes of $u$ in $G'$, *ties broken lexicographically [13] using the unique identifiers of the nodes*.

**An example.** We explain the idea of PSPT using an example (see Figure 1). Suppose we want to compute PSPT of size 5 for each node. We first remove all degree-1 nodes from the network, namely, nodes $\{3, 7, 8, 11, 12, 13, 14, 16\}$. Now, let us construct the PSPT of node 1. Among the remaining nodes, the nodes at distance 1 from node 1 are $\{1, 2, 4, 5, 6, 9\}$. By breaking ties lexicographically, we get that the PSPT of node 1 is given by $\{1, 2, 4, 5, 6\}$ (node 9 is lexicographically larger than the other nodes). This example shows several interesting ideas. First, it may be the case that all the nodes at a specific distance may be contained in the PSPT (all nodes within distance 1 of node 15); on the other hand, it may be the case that the PSPT of a node may not even include all its immediate neighbors (node 1, for instance). Finally, we note that for different nodes, the PSPT may expand to different distances (distance 2 for node 10 while only distance 1 for node 15). **We remark that the network in the example has unit weight edges only for simplicity; our definition of PSPTs and the following discussion does not make any assumption on edge weights.**

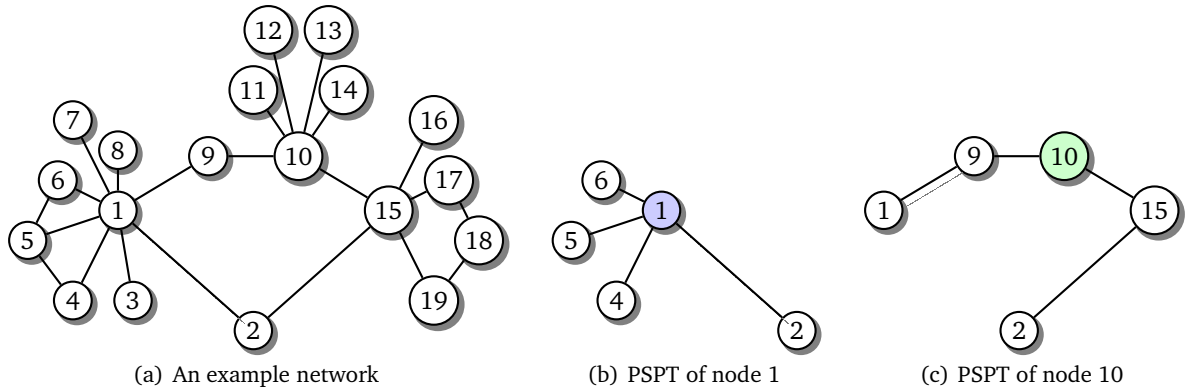

(a) An example network  (b) PSPT of node 1  (c) PSPT of node 10

**Figure 1:** An example to explain the idea of node PSPTs. Here, we construct PSPT of size 5 **for each node.**

## 2.2 The Algorithm

During the preprocessing phase, ASAP computes a data structure that is used to quickly compute paths during the query phase. We start by describing the data structure. ASAP computes and stores three pieces of information during the preprocessing phase:

- for each degree-1 node, the identifier of its (only) neighbor and the distance to this neighbor;
- for each node $u$ of degree greater than 1, the identifier of and the distance to each node $w$ in the PSPT of $u$ of size[3] $4\sqrt{n}$.
- for each node $u$ of degree greater than 1 and for each node $w$ in the PSPT of $u$, the identifier of the first node along the shortest path from $w$ to $u$.

Using the above three pieces of information, ASAP computes the shortest path between any pair of nodes $s$ and $t$ in two steps (see Algorithm 1). Assume that both nodes are of degree greater than 1 and that the two PSPTs intersect along the shortest path. Then, the first step of the algorithm (lines 6–10) finds the node $w_0$ along the shortest path that is contained in both the PSPTs — it iterates through each node $w$ in the PSPT of $s$ and checks if $w$ is contained in the PSPT of $t$; if it does, then the sum of distance from $s$ to $w$ and from $t$ to $w$ is a candidate shortest distance. The node $w$ that corresponds to the minimum of all the candidate distances is in fact node $w_0$.

The second step of the algorithm (lines 11–13) then computes the subpath $w_0 \rightsquigarrow s$ by following the series of next-hops starting $w_0$ until $s$ and the subpath $w_0 \rightsquigarrow t$ by following the series of next-hops starting $w_0$ until $t$ (note that both of these subpaths are completely contained within the data structure constructed during the preprocessing phase). The path from $s$ to $t$ is then returned by concatenating the two subpaths.

[3]The reasons for constructing PSPTs of this specific size are discussed in the next section

We now resolve the two assumptions made in the above description. First, if either of the nodes has degree 1, we replace the node by its neighbor and add the corresponding distance in the result (lines 1–4). Regarding the second assumption, we consider two cases. First, when the PSPT of the two nodes intersect but not along the shortest path, ASAP returns a path that is at most 1 hop longer than the actual shortest path (see Appendix). Second, when the PSPT of the two nodes do not intersect at all, current implementation of ASAP simply runs a bidirectional shortest path algorithm. As we will show in the next section, the latter two cases occur with an extremely low probability.

---

**Algorithm 1** QuerySP$(s, t)$ — algorithm for computing the exact distance between nodes $s$ and $t$. Let $N(u)$ denote the set of neighbors of node $u$ and let $d(u, v)$ denote the exact distance between nodes $u$ and $v$.

---
1: If $|N(s)| = 1$
2:     return $d(s, N(s)) +$ QuerySP$(N(s), t)$
3: If $|N(t)| = 1$
4:     return $d(t, N(t)) +$ QuerySP$(s, N(t))$
5: $\delta \leftarrow \infty$;     $w_0 \leftarrow \emptyset$
6: For each $w$ in PSPT of $s$
7:     If $w$ in PSPT of $t$
8:         If $d(s, w) + d(t, w) < \delta$
9:             $\delta \leftarrow d(s, w) + d(t, w)$
10:            $w_0 \leftarrow w$
11: If $w_0 \neq \emptyset$
12:     Compute path $s \rightsquigarrow w_0$ and path $w_0 \rightsquigarrow t$
13:     Return path $s \rightsquigarrow w_0 \rightsquigarrow t$
14: Else
15:     Run a bidirectional shortest path algorithm

---

In §2.4, we will discuss an extension to Algorithm 1 that allows computing multiple paths between a given pair of nodes. We first discuss an efficient implementation of ASAP that requires lower latency and memory when compared to alternative implementations.

## 2.3 An Efficient Implementation

In this subsection, we describe a low-memory, low-latency implementation for Algorithm 1. **One trivial but inefficient way of storing PSPTs and checking for PSPT intersection is by using hash tables**. In particular, for each node $u$, we construct a hash table with each key being the identifier of a node $w$ in the PSPT of $u$ and the corresponding value being the distance between $u$ and $w$ and the next-hop along the shortest path from $u$ to $w$. The PSPT intersection step in Algorithm 1 can then be trivially implemented using hash table lookups. However, a hash table based implementation is inefficient due to two reasons. First, storing PSPTs using hash tables has a non-trivial memory overhead; our experiments suggest that hash tables can require up to 6–48× more memory when compared to on-disk space requirements. Second, while hash tables have a constant lookup time on an average, the absolute time required for each lookup may be large when compared to, say, comparing two integers.

**For our implementation of ASAP, we used arrays for storing the PSPTs**. In our experiments, an array based implementation required 3-24× less memory and had 4-5× lower latency compared to a hash table based implementation. An array-based implementation of Algorithm 1 is fairly straightforward; we briefly describe it here for sake of completeness.

For each node $u$, an array stores the nodes in the PSPT of $u$ in increasing order of their node identifiers; hence, the node at index $i$ in the array has the $i^{\text{th}}$ smallest identifier among the nodes in the PSPT of $u$. To check for PSPT intersection for a pair of nodes $u$ and $v$, one pointer per array is maintained; each of the pointers is initially set to the first index of the respective array. In each step, the node identifiers corresponding to the two pointers are compared (say $u_i$ and $v_j$). Note that if $u_i > v_j$, none of the nodes $v_k$, $k \leq j$ can have an identifier same as that of $u_i$ (this is where storing nodes in the PSPT in increasing order of identifiers help!); hence, the pointer of $v$ is advanced to $v_{j+1}$. Using the same argument, if $u_i < v_j$, the pointer of $u$ is advanced to $u_{i+1}$. The final case is when $u_i = v_j$. In this case, the node with identifier $u_i = v_j$ lies in both the PSPTs and hence, there is a candidate path of length $d(u, u_i) + d(v, v_j)$ between $u$ and $v$. In this case, the pointers are advanced to $u_{i+1}$ and $v_{j+1}$, respectively. The algorithm terminates when one of the pointers attempt to move beyond the length of the array, and returns the minimum of all candidate path lengths.

To aid path computations, we slightly modify the structure of our array — for each node $u$ and each node $w$ in the PSPT of $u$, the array will now store (in addition to the node identifiers and corresponding distances), the index at which the node identifier of the first hop along the shortest path *from $w$ to $v$* is stored in the array.

## 2.4 Extension for Computing Multiple Paths

We now extend Algorithm 1 for the purpose of computing multiple paths between a given pair of nodes. The high-level idea is to output a path corresponding to each intersection of the two PSPTs while avoiding duplicate paths. To achieve this, we maintain a list of "visited" nodes during the execution of Algorithm 1; these are the nodes that lie along paths that have been output by the algorithm. More specifically, for any pair of nodes $s, t$, upon finding a node $w$ that belongs to the PSPT of both $s$ and $t$, the algorithm first checks if $w$ is marked as visited; if yes, we ignore node $w$. If not, we compute subpaths $s \rightsquigarrow w$ and $w \rightsquigarrow t$ and mark each node along these two subpaths as visited and outputs the path $s \rightsquigarrow w \rightsquigarrow t$, as earlier. It is easy to see that by maintaining such list of visited nodes, the algorithm never outputs duplicate paths.

## 3. EVALUATION

In this section, we evaluate the performance of ASAP over several real-world datasets. We start by describing the datasets and experimental setup (§3.1). We then discuss several properties of node PSPTs (§3.2). Finally, we discuss the performance of ASAP (§3.3).

### 3.1 Datasets and Experimental Setup

The datasets used in our experiments are shown in Table 1. The DBLP dataset is from [7]; the LiveJournal dataset is from [21] and the rest of the datasets are from [15].

**Table 1: Social network datasets used in evaluation.**

| Topologies | # Nodes (Million) | % Nodes with degree $\leq 1$ | # Edges (Million) |
|---|---|---|---|
| DBLP | 0.71 | 13.77% | 2.51 |
| Flickr | 1.72 | 50.95% | 15.56 |
| Orkut | 3.07 | 2.21% | 117.19 |
| LiveJournal | 4.85 | 21.83% | 42.85 |

For each dataset, we sampled 1000 nodes uniform randomly per experiment and repeated the experiment 10 times. The results presented below are, hence, for 10,000 unbiased samples for nodes and 10 million unbiased samples for node pairs.

### 3.2 Properties of PSPTs

We start by empirically studying several interesting properties of node PSPTs with the goal of explaining our specific definition of node PSPTs and of choosing the size of node PSPTs to be $4\sqrt{n}$. To do so, for each dataset, we constructed node PSPTs of size $\alpha \cdot \sqrt{n}$ for $\alpha$ varying from 1/16 to 32 in steps of multiplicative factor 2.

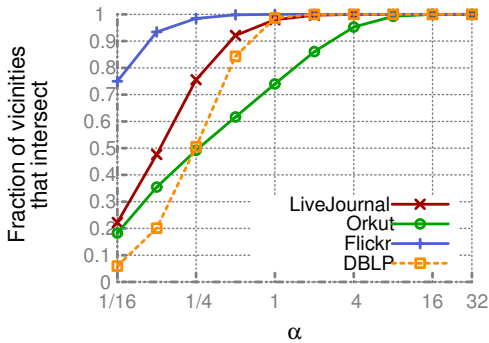 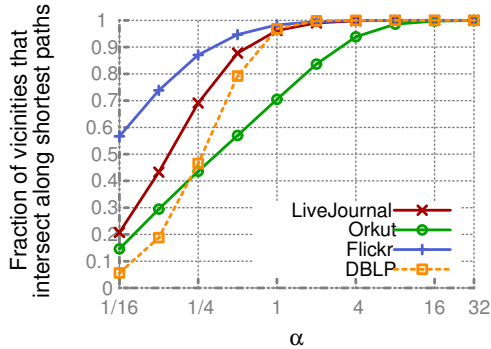

(a) For PSPTs of size $4\sqrt{n}$ and larger, any pair of PSPTs intersects with extremely high probability.

(b) For PSPTs of size $4\sqrt{n}$ and larger, PSPTs almost always intersect along the shortest path.

**Figure 2: Fraction of PSPTs of size $\alpha\sqrt{n}$ that intersect: (a) overall; (b) along the shortest path. For precise values for $\alpha = 2, 4$ and $8$, see Table 2.**

### 3.2.1 PSPT intersection

ASAP builds upon the idea of PSPT intersection to quickly compute shortest paths. Recall that we say that **the PSPTs of a pair of nodes $s, t$ intersect** if there is a node $w$ that is contained in both the PSPT of $s$ and the PSPT of $t$. We evaluate, using the setup described in §3.1, the fraction of pairs of nodes that have intersecting PSPTs in real-world datasets for PSPTs of varying size. Figure 2(a) and Table 2 show the variation of fraction of PSPT intersections with size of the varying PSPTs. Note that (with the exception of the Orkut network) for PSPTs of size $4\sqrt{n}$ and larger, the PSPTs of any two randomly selected nodes intersect with an extremely high probability (more than 0.9998). In fact, for all datasets, PSPTs of size $16\sqrt{n}$ intersect for *each* source-destination pair. We note that the Orkut network has a significantly different structure — it has an extremely high average degree (up to 11× larger than other networks), has very few degree-1 nodes — and yet, shows trends similar to other networks.

### 3.2.2 PSPT intersection along shortest paths

Social networks exhibit a structure much stronger than a large fraction of PSPTs merely intersecting. In particular, empirically, for PSPTs of size $4\sqrt{n}$ and larger, not only do the PSPTs of almost all pairs of nodes intersect, the intersection occurs along the shortest path.

It is easy to see that our definition of the PSPT of a node, due to use of tie-breaking, does not guarantee that for any given pair of nodes, the intersection of PSPTs, if any, occurs along the shortest path. Although our definition of a node PSPT does not guarantee intersection along the shortest path, real-world social networks do exhibit this property for PSPTs of size $4\sqrt{n}$ and larger. Figure 2(b) shows that for PSPTs of size $4\sqrt{n}$, most pairs of nodes have PSPTs intersecting along the shortest path. More interestingly, comparing results of Figure 2(a) and Figure 2(b) (also see Table 2), we note that for PSPTs of size $4\sqrt{n}$ and larger, whenever the PSPTs intersect, they almost always intersect along the shortest path.

For node pairs whose PSPTs intersect but not along the shortest path, we will prove later that the length of the path via node along which the PSPTs intersect is "not too long" when compared to the shortest path. In addition, a non-trivial lower bound can be proved on the distance between node pairs whose PSPTs do not intersect. These observations may be interesting for applications that do not necessarily require computing shortest paths and that require computing shortest path only for pairs of nodes that are "close enough".

### 3.2.3 Benefits of Consistent Tie Breaking

Recall that our definition of node PSPT (§2.1) requires that ties be broken lexicographically using the unique node identifiers. We now elaborate on the significance of this tie breaking scheme. Figure 3 compares the performance of our tie breaking scheme with that of an arbitrary tie breaking scheme as in standard implementations of shortest path algorithms; for our experiments, we used the implementation provided by the Lemon graph library [12].

We observe that when the PSPT sizes are rather small, consistent tie breaking can significantly increase the fraction of PSPTs that intersect along the shortest path. For the LiveJournal network and for PSPTs of size $\sqrt{n}/4$, for instance, consistent tie breaking has 57% more PSPT intersections along the shortest path when compared to arbitrary tie breaking. For moderate PSPT size (those of our interests), the consistent tie breaking has smaller but noticeable effect — for PSPTs of size $4\sqrt{n}$, consistent tie breaking leads to an additional fraction 0.24 PSPT intersections along the shortest path for the Orkut network.

**Table 2: Precise numbers (approximated to four decimal places) for Figure 2 for PSPTs of size $\alpha\sqrt{n}$ for $\alpha = 2, 4$ and 8. For PSPTs of size $4\sqrt{n}$ and larger, almost all PSPT pairs intersect along the shortest path.**

| Dataset | Fraction of interesting PSPTs | | | | | |
|---|---|---|---|---|---|---|
| | $\alpha = 2$ | | $\alpha = 4$ | | $\alpha = 8$ | |
| | total | along shortest path | total | along shortest path | total | along shortest path |
| DBLP | 0.9999 | 0.9986 | 1 | 1.0000 | 1 | 1 |
| Flickr | 1 | 0.9951 | 1 | 0.9993 | 1 | 1.0000 |
| Orkut | 0.8611 | 0.8366 | 0.9530 | 0.9386 | 0.9927 | 0.9859 |
| LiveJournal | 0.9967 | 0.9905 | 0.9998 | 0.9983 | 1.0000 | 0.9998 |

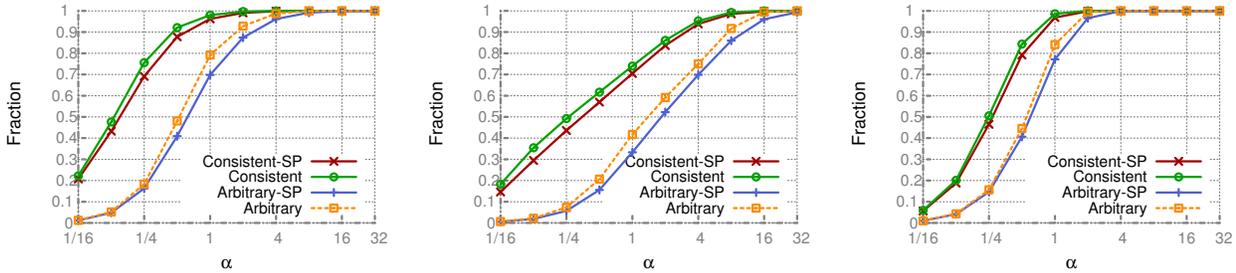

**Figure 3: Comparison of the tie breaking scheme of §2.1 to arbitrary tie breaking (see §3.2.3) for LiveJournal (left), Orkut (middle) and DBLP (right). Consistent-SP and Consistent are for PSPT intersections along shortest path and overall for the tie breaking scheme of §2.1; Arbitrary-SP and Arbitrary are corresponding ones for the arbitrary tie breaking scheme.**

### 3.3 Performance of ASAP

Finally, we discuss the performance of ASAP in terms of preprocessing time, memory requirements, accuracy of computing paths and most importantly, the time taken to compute paths. We discuss these for three specific points of interest — $\alpha = 2, 4$ and 8 (that is, PSPTs of size $2\sqrt{n}, 4\sqrt{n}$ and $8\sqrt{n}$, respectively); these are the values which provide the most interesting trade-offs between memory, latency and accuracy for ASAP and are of practical interest.

#### 3.3.1 Preprocessing and Memory

We start by evaluation ASAP in terms of the time taken to construct the data structure and the memory requirements of the resulting data structure. Note that the preprocessing and memory requirements of ASAP are independent of whether one wants to compute a single shortest path or multiple short paths between a given pair of nodes.

**Preprocessing time.** The results on time taken to construct our data structure for various networks are shown in Table 3. Note that, as expected, the preprocessing time increases with the size of the PSPTs, the size of the network and more importantly, the average degree of the network. **We observe that it is rather easy to distribute the computations required for construct-** ing the data structure across multiple machines — each machine can compute the PSPT for a fraction of nodes in the network. For instance, using just 10 dual-core machines, we can construct the data structure for the LiveJournal and the Orkut networks in less than 8 hours and 13 hours, respectively. This is comparable or faster than the preprocessing time of recent shortest path computation heuristics [4, 26] and significantly faster than techniques that allow computing approximate distances and paths (based on the evaluations in [11]). However, the former set of techniques are limited to computing a single path between any given pair of nodes, have higher latency compared to ASAP and do not admit efficient distributed implementation.

**Memory requirements.** ASAP requires storing, for each node with degree greater than 1, an array with $\alpha\sqrt{n}$ entries. Table 3 shows the average memory requirements per node (on disk and in-memory) for ASAP (if only distances need be retrieved, the memory requirements reduce by roughly 33%). We note that the memory requirements of ASAP, although far from ideal, are much lower than the data structures usually maintained by social networks for answering various user queries. Alternative shortest path computation techniques [4, 26] require slightly lower memory in practice but, unlike ASAP, do not provide any guarantees. In addition, unlike ASAP, these techniques will require significantly higher memory for computing multiple paths between users. Moreover, some of the previous techniques [26] require a hash table based implementation and hence, have extremely high overhead if the data structure is stored in memory; ASAP, on the other hand, employs an array based implementation and hence, has much lower overhead. Our own experiments suggest that arrays require 3-24× less memory than hash tables; hence, an in-memory implementation of ASAP would require memory comparable to that of techniques.

**Table 3: Average preprocessing time & memory requirements for ASAP (approximated to one decimal place).**

| Dataset | Preprocessing time (ms per node) | | | Memory Requirements (kB per node on disk) | | | Memory Requirements (kB per node in-memory) | | |
|---|---|---|---|---|---|---|---|---|---|
| | $\alpha=2$ | $\alpha=4$ | $\alpha=8$ | $\alpha=2$ | $\alpha=4$ | $\alpha=8$ | $\alpha=2$ | $\alpha=4$ | $\alpha=8$ |
| DBLP | 6.9 | 13.6 | 28.2 | 10.4 | 20.9 | 41.9 | 23.7 | 35.4 | 58.6 |
| Flickr | 75.8 | 149.9 | 278.7 | 8.9 | 18.2 | 37.4 | 22.1 | 32.6 | 53.5 |
| Orkut | 130.7 | 298.3 | 638.3 | 29.4 | 58.8 | 117.8 | 39.4 | 66.9 | 121.4 |
| LiveJournal | 56.0 | 113.2 | 237.1 | 28.4 | 57.0 | 114.7 | 39.6 | 67.4 | 122.8 |

### 3.3.2 Accuracy

In terms of accuracy, we make three observations. First, since Algorithm 1 iterates through all the nodes in the PSPT of the source to check for PSPT intersection, ASAP returns the shortest path as long as the PSPTs intersect along the shortest path. From Table 2, this happens for 99.83% of the node pairs for PSPTs of size $4\sqrt{n}$; of course, a higher accuracy of 99.98% can be achieved by using PSPTs of size $8\sqrt{n}$. Second, out of the remaining 0.171% of the node pairs, ASAP returns at least one path for 0.150% of the pairs since their PSPTs intersect; for these node pairs, ASAP provides the following guarantee (see Theorem 1 in the Appendix for the proof): if the distance between the nodes is $d(s,t)$, the distance returned by the algorithm is $d(s,t)+W_{max}$, where $W_{max}$ is the weight of the heaviest edge incident on nodes in the PSPT of the source (for networks modeled as unweighted graphs, $W_{max}=1$). Finally, for the remaining 0.021% of the node pairs, it is possible to combine ASAP with those for computing exact [9, 10] paths; however, it may just be easier to just store shortest paths between such a small fraction of node pairs (as and when they are computed).

### 3.3.3 Query latency

Finally, we discuss the results for query time. Our implementation stores the PSPTs of nodes in-memory using a standard C++ array implementation. The implementation runs on a single core of a Core i7-980X, 3.33 GHz processor running Ubuntu 12.10 with Linux kernel 3.5.0-19. Table 4 compares the query time of ASAP for shortest path computation (for $\alpha=2,4$ and 8) with that of an optimized implementation of breadth-first search algorithm and bidirectional breadth-first search algorithm [10] using the experimental setup of §3.1. Note that we compare the performance of ASAP with breadth-first search simply to demonstrate that *even* for unweighted networks, ASAP provides significant speedups; the relative performance of ASAP will be much better in comparison with a shortest path algorithm for weighted networks since the query time of ASAP is independent of whether the network is weighted or not.

We make several observations. First, rather surprisingly, ASAP is at least 4-5 orders of magnitude faster than the current fastest known technique for computing paths of extremely low error (that is, error less than 10%; the current fastest implementation [11] for computing low error paths requires at least 1090 ms and can be up to 2751 ms for the Orkut network). Second, the latency of ASAP for single path computation is $2-5\times$ lower than the techniques that compute the exact shortest path [4, 26]. Note, however, that ASAP achieves this speed up at the cost of slight loss in accuracy; we believe that such loss in accuracy is completely acceptable for most real-world applications as long as we achieve a speed up for most of the queries.

Not only does ASAP require lower latency for single shortest path computation, its most significant advantage is that it enables computing a large number of paths between any given node pair in less than 100 *microseconds* (see Table 5). Consider the LiveJournal network for instance. ASAP computes 453 paths, on an average, between a given pair of nodes in roughly 99 microseconds, hence enabling a plethora of new social network applications. We are not aware of any other technique that can compute multiple paths between node pairs in time comparable to ASAP.

## 4. A DISTRIBUTED IMPLEMENTATION

ASAP, as presented in §2 computes the shortest path between a given node pair in tens of microseconds. In this section, we show how to implement ASAP in a distributed fashion. This enables ASAP to answer batch shortest path queries without replicating the entire data structure along multiple machines which may be useful for applications with high workload. We will also discuss how to exploit the functionalities offered by distributed programming frameworks like MapReduce [8] and Pregel [14] for an efficient distributed implementation of ASAP.

Table 4: Query time results (approximated to three decimal places) for ASAP.

| Dataset | Our technique Time (in $\mu s$) | | | BFS Time (in ms) | Bidirectional BFS Time (in ms) | Speed-up (compared to Bidirectional BFS) |
|---|---|---|---|---|---|---|
| | $\alpha = 2$ | $\alpha = 4$ | $\alpha = 8$ | | | |
| DBLP | 9.721 | 20.325 | 41.827 | 327.2 | 18.614 | $445 - 1915\times$ |
| Flickr | 12.967 | 26.474 | 51.974 | 2090.2 | 83.956 | $1615 - 6475\times$ |
| Orkut | 15.686 | 31.756 | 64.968 | 28678.5 | 760.987 | $11713 - 48514\times$ |
| LiveJournal | 24.072 | 48.938 | 100.197 | 6887.2 | 156.443 | $1561 - 6499\times$ |

Table 5: Results on query time and number of paths for computing multiple paths using ASAP.

| Dataset | $\alpha = 2$ | | $\alpha = 4$ | | $\alpha = 8$ | |
|---|---|---|---|---|---|---|
| | Time ($\mu s$) | #Paths | Time ($\mu s$) | #Paths | Time ($\mu s$) | #Paths |
| DBLP | 15.080 | 62 | 31.568 | 173 | 68.841 | 416 |
| Flickr | 23.672 | 276 | 51.974 | 523 | 83.956 | 1762 |
| Orkut | 22.250 | 81 | 64.968 | 237 | 760.987 | 817 |
| LiveJournal | 34.462 | 115 | 99.197 | 453 | 156.443 | 1141 |

Recall that the data structure of §2 stores, for each node $u$ in the network[4], the exact distance to each node in the PSPT of $u$; in other words, the data structure stores $\alpha\sqrt{n}$ triplets of the form $\langle u, (w, d(u,w))\rangle$, each corresponding to some node $w$ in the PSPT of $u$. In the following description, we assume that each node $u$ is assigned a machine in the cluster (for instance, using a hash function) and all the triplets corresponding to $u$ are stored on that machine; it is rather trivial to extend ASAP to the case when the triplets for a single node $u$ are split across machines.

A distributed implementation of ASAP is formally described in Algorithm 2. We explain the algorithm for a particular pair of nodes $s$ and $t$. We start the query process by sending the query to the machines that store the triplets for nodes $s$ and $t$. In the first step, the machine storing triplets for node $s$ outputs, for each node $w$ in the PSPT of $s$, a $\langle key, value\rangle$ pair with $w$ being the key and with $(s, d(s,w))$ being the value; we denote this $\langle key, value\rangle$ pair as $\langle w; (s, d(s,w))\rangle$. The machine storing triplets for node $t$ does the same. In the next step, **the algorithm implements PSPT intersection in a distributed fashion**. Specifically, each distinct key is assigned to one machine and all values associated

---
[4]In this section, we assume that all nodes are of degree greater than 1.

**Algorithm 2** A distributed implementation of ASAP; the algorithm computes the shortest distance between all node pairs in a set $Q$.

1: STEP 1 (AT EACH MACHINE):
2: **Input:** triplets for a subset of nodes $S \subset V$
3: For each node $u \in S \cap Q$
4:     For each $w$ in PSPT of $u$
5:         Output $\langle key, value\rangle = \langle w; (u, d(u,w))\rangle$
6:
7: STEP 2 (AT MACHINE ASSIGNED KEY $w$):
8: **Input:** All $\langle key, value\rangle$ pairs with $w$ as the key
9: For each pair of values $(u, d(u,w))$ and $(v, d(v,w))$
10:    Output $\langle key, value\rangle = \langle (u,v); d(u,w)+d(v,w)\rangle$
11:
12: STEP 3 (AT MACHINE ASSIGNED KEY $(u,v)$):
13: **Input:** All $\langle key, value\rangle$ pairs with $(u,v)$ as the key
14: Output the minimum of all the values received with that key (from any machine) are transferred to that machine. Note that for any key $w$, if the machine assigned key $w$ receives two values corresponding to nodes $s$ and $t$, then node $w$ must belong to the PSPTs of both $s$ and $t$ and hence in the intersection of the two PSPTs; hence, there must be a candidate path of length $d(s,w) + d(t,w)$ between $s$ and $t$ — all such candidate paths constitute the output of the second step. As long as the PSPTs intersect along the shortest path, one of these paths (precisely, the path of shortest length) must be the shortest path between $s$ and $t$; the final step computes this path by finding the minimum over all the paths output by machines in the second step.

Table 6: Per-query latency in microseconds amortized over all node pairs using the experimental setup of §3.1 and an external memory (data residing on HDFS and read by mappers) MapReduce implementation.

| Dataset | 20 Mappers & Reducers | | | 40 Mappers & Reducers | | | 80 Mappers & Reducers | | |
|---|---|---|---|---|---|---|---|---|---|
| | Map | Shuffle +Reduce | Total | Map | Shuffle +Reduce | Total | Map | Shuffle +Reduce | Total |
| DBLP | 33.574 | 34.866 | 68.440 | 19.370 | 25.826 | 45.196 | 11.622 | 19.370 | 30.992 |
| Flickr | 140.051 | 225.640 | 365.690 | 73.916 | 143.941 | 217.857 | 38.903 | 105.038 | 143.941 |
| Orkut | 134.462 | 53.785 | 188.247 | 68.725 | 29.880 | 98.605 | 35.856 | 20.916 | 56.772 |
| LiveJournal | 255.796 | 99.561 | 355.357 | 128.664 | 59.737 | 188.401 | 67.395 | 41.356 | 108.751 |

**Extension for retrieving shortest paths.** Let $P(u, v)$ denote the shortest path between any pair of nodes $u$ and $v$. To extend Algorithm 2 to retrieve the shortest path, we use the trick from §2.4 that allows computing the path from $s$ to any node $w$ in the PSPT of $s$. Algorithm 2 can then be modified to return the corresponding paths by simply appending the path information in Step 1. In particular, rather than having values of the form $\langle w; (s, d(s, w))\rangle$, we use values of the form $\langle w; (s, P(s, w), d(s, w))\rangle$. The machines in Step 2 simply concatenate the paths $P(s, w)$ and $P(t, w)$ to return the corresponding path $P(s, t)$.

**Implementing on MapReduce.** We now show how to implement Algorithm 2 on MapReduce using two rounds of operations. The first and the second steps of the algorithm form the Map and the Reduce steps of the first round. The outputs of the second step can be written to the Hadoop distributed file system (HDFS) and can be fed to the mapper in the next round. To implement our algorithm, the mapper of the second round will be an identity function — it simply outputs all $\langle$key, value$\rangle$ pairs as read; finally, the step three forms the reducer step of the second round.

**Memory and Bandwidth requirements.** Let $p$ be the total number of machines in the cluster. We now argue that ASAP requires each machine to store at most $\alpha n \sqrt{n}/p$ entries. We start by noting that since the data structure is distributed across the set of machines, the memory required at any single machine in the first step is simply a factor $1/p$ when compared to a single machine implementation. In the second step, each node $w$ can be in the PSPT of at most $n$ nodes, and hence, each machine requires storing $n$ entries. In the last step, each machine requires storing exactly one entry (to keep track of the shortest path seen so far). For the bandwidth requirements, we note that ASAP transfers $\alpha \sqrt{n}$ entries corresponding to each node that participates in the query.

**Implementation results.** Finally, we evaluate the performance of distributed ASAP using an external memory MapReduce implementation, that is, the data is residing on the HDFS and is read when queries are initiated. The implementation runs on a 64 node MapReduce cluster with Hadoop version 0.19.1; each node in the cluster supports up to 4 mappers and 4 reducers. The evaluation results are shown in Table 6.

We make several observations. First, we note that since distributed ASAP requires significantly less memory when compared to a single machine implementation of ASAP, it is entirely feasible to do an in-memory implementation of distributed ASAP (our cluster does not provide support for this); for instance, for the LiveJournal network, using 40 machines require roughly 6.5 GB of memory which is feasible for most modern desktops. With such an in-memory implementation, the corresponding query latency will simply be the time consumed in the shuffle and reduce phase and hence, less than 100 microseconds for most networks. Second, even with an external memory implementation, the amortized query time is less than 366 microseconds and most of this is spent reading the input data from HDFS[5]. Finally, we note that the amortized query latency (and memory requirements!) of distributed ASAP reduce almost linearly with increase in the number of machines, which is a highly desirable property of distributed implementations.

## 5. RELATED WORK

Our goals are related to two key areas of related work: algorithms and heuristics for computing shortest paths and for computing approximately shortest paths.

**TEDI and Pruned Landmark Labeling.** TEDI [26] is one of the closest work related to ASAP. Independent to our work, a recent paper named Pruned Landmark

---

[5] We note that the mappers in our distributed algorithm perform extremely simple tasks; hence, the extremely high time of mapper operations (255 microseconds for the LiveJournal network with 20 mappers and reducers) is due to slow hard disks and issues with the filesystem. The same files can be read on a single machine using our Ubuntu based implementation in amortized time of less than 12 microseconds.

Labeling [4] also computes shortest paths in large social networks. We compare the performance of ASAP with [4, 26] in terms of preprocessing time, memory, accuracy and latency. The preprocessing time of [26] is significantly larger than that of ASAP; [4], on the other hand, requires lower preprocessing time. However, as discussed in §3, the preprocessing stage can be easily parallelized across multiple machines and hence, is not really[6] a bottleneck. In terms of memory and accuracy, the focus of [4, 26] was on providing 100% accuracy (shortest paths for all node pairs) without providing any bounds on memory requirements; although their memory footprint is lower than that of ASAP. ASAP, on the other hand, provides guarantees of the memory requirements with slight loss in accuracy. Either of these trade-offs may be interesting depending on the application. **The main advantage of ASAP is its lower latency, its ability to compute multiple paths and its ability to answer batch queries using a simple distributed implementation**. None of [4, 26] achieve any of the last two properties.

A workshop version of this paper [1] allowed quickly computing shortest paths on social networks. However, [1] used a different definition of node PSPTs and a hash table based implementation, resulting in higher memory requirements and roughly an order of magnitude higher latency. In addition, [1] did not allow computing multiple paths and did not have a distributed implementation.

**Shortest path algorithms and heuristics.** Heuristics like $A^\star$ search [9, 10] and bidirectional search [10] have been proposed to overcome the latency problems with traditional algorithms for computing shortest paths. The approaches in [9, 10], although useful in reducing the query time, still require running a (modified) shortest path algorithm for each query and do not meet the latency requirements. For instance, the experimental results in §2 shows that bidirectional search can take hundreds of milliseconds to compute the shortest paths even on moderate size networks.

In comparison to [9, 10], our contributions are twofold: first, we show that empirically, in social networks, vicinities of size $4\sqrt{n}$ nearly always intersect (heuristics in [9, 10] could also exploit this); and second, we argue that the vicinities being a small fraction of the entire network, storing and checking intersection quickly is feasible. This should be substantially faster than traditional bidirectional search [10] because it is just a series of hash table look-ups in a relatively compact data structure with one element per vicinity node — as opposed to running a shortest path algorithm that would require priority queue operations, and may even

---

[6]Note that it is not clear how to parallelize the preprocessing stage of [26].

explore a large fraction of the entire network.

**Approximation algorithms.** Arguing that the above heuristics [9, 10] are unlikely to meet the stringent latency requirements of social network applications, [11, 18, 19, 24, 27, 28] focus on computing *approximate* distances and paths. The body of work can be broadly characterized into two categories.

The first category uses techniques from graph embedding literature [27, 28]. The main advantage of these schemes is their low memory footprint; however, these schemes often compute paths of high worst-case stretch (providing a guarantee of $\log(n)$ stretch for a network with $n$ nodes) [27, 28], are often not able to compute shortest paths [27], and require reconstructing the entire data structure from scratch in case of network updates [27, 28]. ASAP, on the other hand provides latency similar to the above techniques while providing the benefits of computing *shortest distances and paths* for most source-destination pairs and efficient update of the data structure upon network updates.

The second category uses techniques from distance oracle literature [11, 18, 19, 24]. In comparison to these techniques, ASAP differs in several aspects. First, the above techniques are primarily modifications or heuristic improvements on results from theoretical computer science [23]; these results are now known to be far from optimal for real-world networks [2, 3, 17] which ASAP borrows ideas from. Second, techniques in this category that have lowest latency [19] return paths that have high absolute error (more than 3 hops on an average, even on small networks); in comparison, ASAP computes shortest paths between almost all source-destination pairs. On the other hand, techniques that provide significantly better accuracy require 4-5 orders of magnitude higher query time when compared to ASAP [11, 24]. Third, similar to graph embedding based techniques, some of these techniques [18] are unable to compute the actual paths. Finally, distance oracle based techniques are known to not admit efficient algorithms for updating the data structure requiring a large number of single-source shortest path computations upon each update and each such computation takes time in the order of tens to hundreds of seconds; in contract, dynamic-ASAP can update the data structure in less than half a second.

## 6. DISCUSSION

In this section, we highlight some limitations of ASAP in the form of the most frequently asked questions.

**Does ASAP work on all input networks?** In short, no. There are two main ideas used in design of ASAP— (1) a definition of node PSPTs such that most pairs of PSPTs intersect along the shortest path[7]; and (2) an efficient implementation that exploits the first idea to quickly compute shortest paths. Our focus in this paper was on social networks, where we have shown that the structure of social networks enable a suitable definition of node PSPTs. Indeed, there are networks (line networks and grid networks, for instance) for which our definition of node PSPTs will not adhere to the first idea above. However, for such networks, there is another definition of node PSPTs that provably guarantees that all pairs of PSPTs intersect [6] and node PSPTs can be used to compute the shortest path. It remains an interesting (and apparently, unresolved) question to identify networks for which no suitable definition of PSPTs exist to guarantee that most pairs of PSPTs intersect and paths can be efficiently computed. In addition, ASAP is currently designed for networks modeled as undirected graphs; extending ASAP to handle directed networks is an interesting avenue for future work.

**Doesn't ASAP have high memory requirements?** As discussed in §3.3.1, ASAP does require more *disk space* than techniques that compute approximately shortest paths; for instance, the disk space required by ASAP for the Orkut dataset is roughly 172 GB, an order of magnitude larger than the fastest technique for computing approximately shortest paths [11]. We make several remarks assuming that such schemes are implemented in-memory (for external memory implementations, disk space is not really a major concern). First, social networks typically maintain much larger datasets for answering various user queries; hence, given the low latency and high accuracy of ASAP, such memory requirements are quite acceptable. Second, the techniques in [11] naturally require a hash table based implementation that has significantly higher overhead when compared to ASAP's array-based implementation; our own experiments suggest that a hash table based implementation requires 3-24× higher memory when compared to an array-based implementation. Hence, for an in-memory implementation, the memory requirements of ASAP will be comparable to that of techniques that compute approximately shortest paths. Finally, if memory were really a bottleneck, one could use dynamic-ASAP that has extremely low memory requirements (just 3.08 GB of disk space for the Orkut network) and can compute shortest paths at least two orders of magnitude faster than the technique in [11].

## 7. CONCLUSION

In this paper, we have presented ASAP, a system that efficiently computes shortest paths on massive social networks by exploiting their structure. Using evaluations on real-world datasets, we have shown that ASAP computes shortest paths for most node pairs in tens of microseconds even on networks with millions of nodes and edges using reasonable memory requirements. We have also shown that ASAP, unlike any previous technique, allows computing multiple paths between a given node pair in less than hundred microseconds using the same data structure as that for single shortest path computation. Finally, we have also shown that ASAP admits efficient distributed implementation and can compute shortest paths between millions of pairs of nodes in well under a second using an in-memory implementation. We believe that the most interesting related future work is to extend ASAP to compute shortest paths on dynamic graphs (of course, without re-computing the data structure from scratch).

## Acknowledgment

The authors would like to thank Chris Conrad from LinkedIn for providing the motivation for computing multiple shortest paths in large social networks.

---

[7]and of course, allows computing shortest paths using the node PSPTs.

# APPENDIX

THEOREM 1. *For any pair of nodes $s, t$ at distance $d(s, t)$, if the PSPTs of $s$ and $t$ have a non-empty intersection, Algorithm 1 either returns the shortest path or returns a path of length $d(s, t) + W_{\max}$, where $W_{\max}$ is the weight of the heaviest edge incident on nodes in the PSPT of $s$.*

PROOF. Let $w$ be the node in the intersection of the PSPTs of $s$ and $t$ that minimizes $d(s, w) + d(t, w)$. If the PSPTs intersect along the shortest path, we get that the returned path is indeed the shortest path since the node $w$ that lies along the shortest path in the intersection minimizes the above expression. Consider the case when $w$ does not lie along the shortest path.

Let $P = (s, v_0, \ldots, v_k, t)$ be the shortest path between $s$ and $t$ and let $i_0 = \max\{i : v_i \in P \cap \Gamma(s)\}$. Note that $v_{i_0} \notin \Gamma(t)$, else by setting $w = v_{i_0}$, we will contradict the assumption that $w$ does not lie along the shortest path between $s$ and $t$. By definition of the vicinity, we have that $d(s, w) \le d(s, v_{i_0}) + W_{\max}$. Furthermore, since $v_{i_0} \notin \Gamma(t)$, we have that $d(t, w) \le d(t, v_{i_0})$. Hence, we get that the returned distance is at most $d(s, w) + d(t, w) \le d(s, v_{i_0}) + W_{\max} + d(t, v_{i_0}) = d(s, t) + W_{\max}$, leading to the proof. □